\documentstyle[12pt]{article}
\newfont{\ffont}{msym10}                        
\newcommand{\beq}{\begin{equation}}             
\newcommand{\eeq}{\end{equation}}               
\newcommand{\bqry}{\begin{eqnarray}}            
\newcommand{\eqry}{\end{eqnarray}}              
\newcommand{\bqryn}{\begin{eqnarray*}}          
\newcommand{\eqryn}{\end{eqnarray*}}            
\newcommand{\preprint}[1]{\begin{table}[t]      
            \begin{flushright}                  
            \begin{large}{#1}\end{large}        
            \end{flushright}                    
            \end{table}}                        
\newcommand{\PD}[2]                             
    {\frac{\partial^{#2}}{\partial #1^{#2}}}    
\catcode`\@=11 \@addtoreset{equation}{section}  
\renewcommand{\theequation}                     
         {\arabic{section}.\arabic{equation}}   
\begin{document}
\preprint{TAUP-2239-95 \\ }
\title{Towards a Realistic Equation of State \\ of Strongly Interacting
Matter}
\author{\\ L. Burakovsky,\thanks {Bitnet: BURAKOV@TAUNIVM.TAU.AC.IL.} \
L.P. Horwitz\thanks{Bitnet: HORWITZ@TAUNIVM.TAU.AC.IL. Also at Department
of Physics, Bar-Ilan University, Ramat-Gan, Israel  } \ \\   \\
School of Physics and Astronomy \\ Raymond and Beverly Sackler
Faculty of Exact Sciences \\ Tel-Aviv University, Tel-Aviv 69978, Israel
\\  \\ and \\  \\ W.C. Schieve\thanks{Bitnet: WCS@MAIL.UTEXAS.EDU.}\ \\
 \\ Ilya Prigogine Center \\ for Studies in Statistical Mechanics, \\
University of Texas at Austin, \\ Austin, TX 78712, USA \\}
\date{ }
\maketitle
\begin{abstract}
We consider a relativistic strongly interacting Bose gas. The interaction
is manifested in the off-shellness of the equilibrium distribution. The
equation of state that we obtain for such a gas has the properties of a
realistic equation of state of strongly interacting matter, i.e., at low
temperature it agrees with the one suggested by Shuryak for hadronic
matter, while at high temperature it represents the equation of state of
an ideal ultrarelativistic Stefan-Boltzmann gas, implying a phase
transition to an effectively weakly interacting phase.
\end{abstract}

\bigskip
\section{Introduction}
Quantum chromodynamics (QCD), the fundamental theory of strong
interactions, can be perturbatively solved only in the region of
asymptotic freedom, i.e., for large momenta of quarks and gluons
\cite{Pol}. At low momenta, quarks and gluons interact strongly and are
confined inside hadrons. In this case, the expansion parameter of
perturbation theory, the running coupling constant $\alpha _s,$ is of the
order of unity, so that perturbative methods are not applicable and one
has to use non-perturbative methods. One such method is the construction
of effective theories to describe the behaviour of hadronic matter, such
as the $\sigma -\omega $ model \cite{Val}. However, one of the most
successful non-perturbative methods is the use of lattice gauge
calculations \cite{Cre}, which are especially suitable for studying
perturbative as well as non-perturbative effects in QCD. The presently
available lattice data mostly concern simulations of $SU(N)$ pure gauge
theory \cite{B,D}, since the treatment of dynamical fermions on a lattice
is difficult \cite{Cre}. Moreover, lattice artefacts are believed to be
well under control only in the pure gauge case \cite{EE}. A striking
feature of lattice simulations of $SU(2)$ and $SU(3)$ pure gauge theory
is a phase transition (of apparently first order for $SU(3)$
\cite{B,D,Ben} and second order for $SU(2)$ theory \cite{Fig}) from a
phase of confined gluons (``glueballs'')
to one of deconfined gluons (``gluon plasma''), leading to a sharp
rise in the energy density of a gluon gas as a function of temperature
at a phase transition temperature $T_c$ \cite{B,D} (Figs. 1,2). In the
$\sigma-\omega $ model, a similar phase transition (or, at least, a rapid
increase of the energy density in a small temperature interval) appears
at small net baryon densities \cite{The} (Fig. 3). This is due to a
strong enhancement of the scalar meson interaction at $T_c\simeq 200$
MeV, leading to a transition from a phase of massive baryons to a phase
of massless baryons. Apparently, a phase transition to a weakly
interacting phase seems to be a fundamental feature of strongly
interacting matter\footnote{It should be, however, mentioned that a phase
transition in the $\sigma -\omega $ model may be an artefact of the
approximations made in the calculation of the thermodynamic quantities in
Fig. 3 (in the mean-field approximation) \cite{HE}.} \cite{Ri}.

To understand the lattice data from a simple physical point of view,
Rischke {\it et al.} \cite{Ri,Ri1} have constructed a phenomenological
model for the gluon plasma, in which gluons with large momenta are
considered as an ideal gas with perturbative corrections of order
$O(\alpha _s),$ while gluons with low momenta are subject to confining
interactions and do not contribute to the energy spectrum of free gluons.
The equation of state for this model, although it quantitatively
reproduces the lattice data for the thermodynamic functions of $SU(3)$
pure gauge theory above the deconfinement transition temperature
(Fig. 2), it has a rather complicated form and is therefore not suitable
for practical use, e.g., in astrophysics (for description of stellar
structure), or in cosmolody (for treatment of a hadron-plasma phase
transition in the early universe). In this paper, we suggest another
equation of state, which reflects the main properties of strongly
interacting matter (i.e., a phase transition to a weakly interacting
phase) and agrees qualitatively with the lattice data for $SU(2)$ pure
gauge theory $both$ above and below the transition temperature, and has a
much simpler form, as compared to that of ref. \cite{Ri}.

First, let us note that in strongly interacting matter, particles
undergoing continual mutual interaction are necessarily off-shell.
Therefore, the effect of strong interaction in such a system may be
represented by the off-shellness of its particles. The equilibrium state
of such a system should be characterized by a well-defined relativistic
mass distribution around the on-shell value. Thus, instead of dealing
with interaction explicitly, we reduce the problem to the description of
the relativistic off-shell ensemble. The role of interaction then
consists in determining the effective thermodynamic parameters
governing the mass distribution in a strongly interacting system.

The physical framework for the description of a relativistic off-shell
ensemble has been established by Horwitz and Piron \cite{HP} as a
manifestly covariant relativistic dynamics whose consistent formulation
is based on the ideas of Fock \cite{Fock} and Stueckelberg \cite{Stu},
in which the four components of energy-momentum are considered as
independent degrees of freedom, permitting fluctuations from the mass
shell. In this framework, the dynamical evolution of a system of $N$
particles, for the classical case, is governed by equations of motion
that are of the form of Hamilton equations for the motion of $N$ $events$
which generate the space-time trajectories (particle world lines) as
functions of a continuous Poincar\'{e}-invariant parameter $\tau $
\cite{HP,Stu}. These events are characterized by their positions $q^\mu =
(t,{\bf q})$ and energy-momenta $p^\mu =(E,{\bf p})$ in an
$8N$-dimensional phase-space. For the quantum case, the system is
characterized by the wave function $\psi _\tau (q_1,q_2,\ldots ,q_N)\in
L^2(R^{4N}),$ with the measure $d^4q_1d^4q_2\cdots d^4q_N\equiv d^{4N}q,$
$(q_i\equiv q_i^\mu ;\;\;\mu =0,1,2,3;\;\;i=1,2,\ldots ,N),$ describing
the distribution of events, which evolves with a generalized
Schr\"{o}dinger equation \cite{HP}. The collection of events (called
``concatenation'' \cite{AHL}) along each world line corresponds to a
{\it particle,} and hence, the evolution of the state of the $N$-event
system describes, {\it a posteriori,} the history in space and time of
an $N$-particle system.

For a system of $N$ interacting events (and hence, particles) one takes
\cite{HP} (we use the system of units in which $\hbar =c=k_B=1;$ we also
use the metric $g^{\mu \nu }=(-,+,+,+))$
\beq
K=\sum _i\frac{p_i^\mu p_{i\mu }}{2M}+V(q_1,q_2,\ldots ,q_N),
\eeq
where $M$ is a given fixed parameter (an intrinsic property of the
particles), with the dimension of mass, taken to be the same for all the
particles of the system. The Hamilton equations are
$$\frac{dq_i^\mu }{d\tau }=\frac{\partial K}{\partial p_{i\mu }}=\frac{p_
i^\mu }{M},$$
\beq
\frac{dp_i^\mu }{d\tau }=-\frac{\partial K}{\partial q_{i\mu }}=-\frac{
\partial V}{\partial q_{i\mu }}.
\eeq
In the quantum theory, the generalized Schr\"{o}dinger equation
\beq
i\frac{\partial }{\partial \tau }\psi _\tau (q_1,q_2,\ldots ,q_N)=K
\psi _\tau (q_1,q_2,\ldots ,q_N)
\eeq
describes the evolution of the $N$-body wave function
$\psi _\tau (q_1,q_2,\ldots ,q_N).$

In the present paper we restrict ourselves to a relativistic Bose gas, in
order to compare the results with experimental data of pure gauge theory
lattice simulations. We show that our results agree with those for
$SU(2)$ pure gauge theory. It should be, however, noted that since the
underlying theory is basically different from QCD, a comparison with the
$SU(2)$ lattice data can only be qualitative. From this point of view,
the similarity is remarkable.

\section{Ideal relativistic Bose gas}
Gibbs ensembles in a manifestly covariant relativistic classical and
quantum mechanics were derived by Horwitz, Schieve and Piron \cite{HSP}.
To describe an ideal gas of events obeying Bose-Einstein statistics in
the grand canonical ensemble, we use the expression for the number of
events found in \cite{HSP} (for our present purposes we assume no
degeneracy),
\beq
N=V^{(4)}\sum _{k^\mu }n_{k^\mu }=
V^{(4)}\sum _{k^\mu }\frac{1}{e^{(E-\mu -\mu _K\frac{m^2}{2M})/T}-1},
\eeq
where $V^{(4)}$ is the system's four-volume and $m^2\equiv -k^2=-k^\mu
k_\mu $ is the variable dynamical mass. Here, in addition to the usual
chemical potential $\mu ,$ there is the mass potential $\mu _K$
corresponding to the Lorentz scalar function $K(p,q)$ (Eq. (1.1)), here
taken in the ideal gas limit, on the $N$-event relativistic phase space;
in order to simplify subsequent considerations, we shall take it to be a
fixed parameter (which determines
an upper bound of the mass distribution in the ensemble we are studying,
as we shall see below). To ensure a positive-definite value for $n_{k^\mu
},$ the number of bosons with four-momentum $k^\mu ,$ we require that
\beq
m-\mu -\mu _K\frac{m^2}{2M}\geq 0.
\eeq
The discriminant of Eq. (2.2) must be nonnegative, which gives
\beq
\mu \leq \frac{M}{2\mu _K}.
\eeq
For such $\mu ,$ (2.2) has the solution
\beq
\frac{M}{\mu _K}\left( 1-\sqrt{1-\frac{2\mu \mu _K}{M}}\right) \leq m\leq
\frac{M}{\mu _K}\left( 1+\sqrt{1-\frac{2\mu \mu _K}{M}}\right) .
\eeq
For small $\mu \mu _K/M,$ the region (2.4) may be approximated by
\beq
\mu \leq m\leq \frac{2M}{\mu _K}.
\eeq
One sees that $\mu _K$ plays a fundamental role in determining an upper
bound of the mass spectrum, in addition to the usual lower bound $m\geq
\mu .$

Replacing the sum over $k^\mu $ (2.1) by an integral, one obtains for
the density of events per unit space-time volume $n\equiv N/V^{(4)}$
\cite{ind},
\beq
n=\frac{1}{4\pi ^3}\int _{m_1}^{m_2}\frac{m^3\;dm\;\sinh ^2\beta \;d
\beta }{e^{(m\cosh \beta -\mu -\mu _K\frac{m^2}{2M})/T}-1},
\eeq
where $m_1$ and $m_2$ are defined in Eq. (2.4), and we have used the
parametrization \cite{HSS} $$\begin{array}{lcl}
p^0 & = & m\cosh \beta , \\
p^1 & = & m\sinh \beta \sin \theta \cos \phi , \\
p^2 & = & m\sinh \beta \sin \theta \sin \phi , \\
p^3 & = & m\sinh \beta \cos \theta ,
\end{array} $$ $$0\leq \theta <\pi ,\;\;\;0\leq \phi <2\pi ,\;\;\;-\infty
<\beta <\infty .$$

In what follows we shall take $\mu \simeq 0$ (as for the case of the
ensemble of gauge bosons). The integral (2.6) is calculated in ref.
\cite{BHS} (in the high-temperature Boltzmann approximation for the
integrand):
\beq
n=\frac{T^4}{4\pi ^3}\left[ 2-x^2K_2(x)\right] ,\;\;\;\;\;x=\frac{
2M}{T\mu _K}\equiv \frac{2m_c}{T},
\eeq
where $K_\nu (z)$ is the Bessel function of the third kind (imaginary
argument), $m_c=M/\mu _K$ is the central value around which the mass of
the particles are distributed, in view of (2.4), and $2m_c$ is the upper
limit of the mass distribution in our case of small $\mu ,$ in view of
(2.5). One calculates the characteristic averages \cite{BHS} to be
\bqry
\langle E\rangle  & = & T\;\frac{8-x^3K_3(x)}{2-x^2K_2(x)}, \\
\langle E^2\rangle  & = & T^2\;\frac{40-x^4K_4(x)+x^3K_3(x)}{
2-x^2K_2(x)}, \\
\langle m^2\rangle  & = & T^2\;\frac{16-x^4K_4(x)+4x^3K_3(x)}{
2-x^2K_2(x)}, \\
\langle {\bf p}^2\rangle  & = & 3T^2\;\frac{8-x^3K_3(x)}{2-x^2K_2(x)}\;=
\;3T\langle E\rangle ,
\eqry
and obtains the following thermodynamic functions (the particle number
density, pressure and energy density) \footnote{If we had not used the
Boltzmann limit for the integrand in (2.6), one would obtain the factors
$\zeta (3)\approx 1.202$ in Eq. (2.12) and $\zeta (4)=\pi ^4/90\approx
1.082$ in Eqs. (2.13),(2.14).}:
\bqry
N_0 & = & \langle J^0\rangle \;=\;\frac{T^3}{\pi ^2}\;\frac{8-x^3K_3(
x)}{x^2}, \\
p & = & \frac{1}{3}\langle T^{ii}\rangle g_{ii}=\;\frac{T^4}{\pi ^2}\;
\frac{8-x^3K_3(x)}{x^2}\;=\;N_0T, \\
\rho  & = & \langle T^{00}\rangle \;=\;\frac{T^4}{\pi ^2}\;\frac{
40-x^4K_4(x)+x^3K_3(x)}{x^2},
\eqry
where $\langle J^\mu \rangle $ and $\langle T^{\mu \nu }\rangle $ are
the average $particle$ four-current and energy-momentum tensor,
respectively, given by \cite{HSS}:
\beq
\langle J^\mu \rangle =\frac{T_{\triangle V}}{M}n\langle p^\mu \rangle ,
\;\;\;\langle T^{\mu \nu }\rangle =
\frac{T_{\triangle V}}{M}n\langle p^\mu p^\nu \rangle .
\eeq
In (2.15), $T_{\triangle V}$ is the average passage interval in $\tau $
for the events which pass through the small (minimal typical) four-volume
$\triangle V$ in the neighborhood of the $R^4$-point; it is related to
a width of the mass distribution around the central value, $\triangle m,$
as follows \cite{BH}:
\beq
T_{\triangle V}\triangle m=2\pi .
\eeq
The expressions (2.13),(2.14) for $p$ and $\rho $ are obtained from Eq.
(2.15) for $\langle T^{\mu \nu }\rangle .$ They are, moreover,
thermodynamically consistent. One may varify easily that the relation
\beq
\rho =T\frac{dp}{dT}-p
\eeq
is satisfied \cite{BHS}.

For low $T,$ Eqs. (2.13),(2.14) reduce, through the asymptotic formula
\cite{AS1}
\beq
K_\nu (z)\sim \sqrt{\frac{\pi }{2z}}e^{-z}\left( 1+\frac{4\nu ^2-1}{8z}+
\ldots \right) ,\;\;\;z>>1,
\eeq
to
\beq
p=\frac{2T^6}{\pi ^2m_c^2},\;\;\;\rho =\frac{10T^6}{\pi ^2m_c^2}=5p,
\eeq
consistent with the ``realistic'' equation of state suggested by
Shuryak for strongly interacting hadronic matter \cite{Shu}.

For high $T,$ we use another asymptotic formula \cite{AS2},
\beq
K_\nu (z)\sim \frac{1}{2}\Gamma (\nu )\left( \frac{z}{2}\right) ^{-\nu }
\left[ 1-\frac{z^2}{4(\nu -1)}+\ldots \right] ,\;\;\;z<<1,
\eeq
and obtain
\bqry
p & = & \frac{T^4}{\pi ^2}\left( 1-\frac{m_c^2}{2T^2}\right) \;=\;
p_{{\rm SB}}\left( 1-\frac{m_c^2}{2T^2}\right) , \\
\rho  & = & \frac{T^4}{\pi ^2}\left( 3-\frac{m_c^2}{2T^2}\right) \;=\;
\rho _{{\rm SB}}\left( 1-\frac{m_c^2}{6T^2}\right) ,
\eqry
where $p_{{\rm SB}}=T^4/\pi ^2$ and $\rho _{{\rm SB}}=3p_{{\rm SB}}$ are
the pressure and energy density of an ideal ultrarelativistic
Stefan-Boltzmann gas. Therefore, as $T\rightarrow \infty ,$ the
thermodynamic functions of the relativistic Bose gas we are considering
become asymptotically those of a Stefan-Boltzmann gas, implying a phase
transition to an effectively weakly interacting phase, from the phase of
strong interactions described by Eq. (2.19).

It follows from Eq. (2.10), via the asymptotic formulas (2.18),(2.20),
that
\beq
\frac{\langle m^2\rangle }{T^2}=\left\{ \begin{array}{ll}
8, & T<<2m_c, \\
2m_c^2/T^2, & T>>2m_c.
\end{array} \right.
\eeq
The dependence of $p/p_{{\rm SB}},$ $\rho /\rho _{{\rm SB}}$ and $\langle
m^2\rangle /T^2$ on temperature are shown in Figs. 4,5. At $T\simeq 0.2\;
m_c$ (corresponding to $z\simeq 0.1$ in Figs. 4,5), there is a smooth
phase transition to a weakly
interacting phase described by Eqs. (2.21),(2.22).

\section{Concluding remarks}
The manifestly covariant framework discussed in the present paper can be
an effective tool in dealing with realistic physical systems. The
equation of state (2.12)-(2.14) obtained in our work reflects the main
properties of strongly interacting matter (i.e., a phase transition to a
weakly interacting phase), and agrees qualitatively
with the lattice data for $SU(2)$ pure gauge theory.

The question naturally arises of why the $SU(2)$ lattice data \cite{Fig}
appear to contain a second order phase transition but the $SU(3)$ data
\cite{B,D} appear to contain first order one. There is no {\it a priori}
difference between $SU(2)$ and $SU(3)$ pure gauge theories from a
statistical mechanical point of view. As shown in ref. \cite{B}, $SU(3)$
pure gauge theory simulations on $24^3\times N_T$ lattices indicate a
rapid rise in $\rho +p$ as a function of temperature which takes place
in the case of $N_T=4,$ reflecting a sharp first order phase transition.
This rise is broadened considerably in the case of $N_T=6,$ with the
slope of the curve diminished by almost a factor 3, indicating a smoother
transition in this case. As remarked by the authors, this softening of
the structure of the transition as $N_T$ is increased may well continue
as the continuum limit is aproached. Thus, in ref. \cite{B} the apparent
first order nature of the transition in the case of $SU(3)$ pure gauge
theory has been called in question. Moreover, there are indications from
lattice QCD calculations that when fermions are included, the phase
transition may be of second or higher order \cite{Ben1}. Altogether,
these observations suggest that a realistic equation of state of strongly
interacting hadronic matter should be expected to contain a second or
higher order phase transition, as reflected by the equation of state
obtained in our work.

As remarked by Ornik and Weiner \cite{OW}, a $single$ equation of state
which, at high temperature, describes a quark-gluon phase and, at low
temperature, a hadronic phase and which contains a phase transition of
either first or higher order provides a more satisfactory theoretical
description than one in which each phase is described by a different
equation of state. In this way, the equation of state that we obtained
can be considered as a candidate for such a realistic equation of state
of strongly interacting matter, in contrast to the equations of state of,
e.g., Rischke {\it et al.} \cite{Ri} and Shuryak \cite{Shu} each of which
describes just one of the phases (above and below the transition,
respectively). The introduction of the quark degrees of freedom in this
equation of state, as well as taking into account an effective
interaction potential in an explicit form in the strongly interacting
phase, and perturbative corrections in the weakly interacting phase,
should enable one to derive an equation of state which we expect to be
more accurate for the description of the phenomena taking place in
strongly interacting hadronic matter. The derivation of such an equation
of state and its possible implications in astrophysics and cosmology
are now being worked out by the authors.

\section*{Acknowledgements}
We wish to thank J. Eisenberg for very valuable discussions, and E.
Eisenberg for his help in drawing the pictures for the present paper.

\newpage
\centerline{FIGURE CAPTIONS}
\bigskip
\bigskip
\bigskip
\bigskip
\hfil\break
Fig. 1. The energy density of $SU(2)$ pure gauge theory on a $10^3\times
3$ lattice as a function of $T/\Lambda _L.$ Taken from ref.
[8].\hfil\break
\hfil\break
\hfil\break
\hfil\break
Fig. 2. Fit of the thermodynamic functions, according to the equation of
state of ref. [10], to the lattice data for
$SU(3)$ pure gauge theory. Taken from ref. [10].\hfil\break
\hfil\break
\hfil\break
\hfil\break
Fig. 3. Phase transition in the $\sigma -\omega $ model. Taken from ref.
[10].\hfil\break
\hfil\break
\hfil\break
\hfil\break
Fig. 4. The pressure and energy density, according to the equation of
state (2.13),(2.14), as functions of $z=T/2m_c.$
\hfil\break
\hfil\break
\hfil\break
\hfil\break
Fig. 5. The average mass squared, according to Eq. (2.10), as a
function of $z=T/2m_c.$
\end{document}